\documentclass[aip,preprint]{revtex4-1}
\usepackage{graphicx}

\usepackage{appendix}
\usepackage{amsmath}
\usepackage{comment}
\usepackage{color}

\draft 

\begin{document}

\title{Simulation of strongly quantum-degenerate uniform electron gas using the pseudo-fermion method}

\author{Yunuo Xiong}
\email{xiongyunuo@hbpu.edu.cn}
\affiliation{Center for Fundamental Physics, Hubei Polytechnic University, Huangshi 435003, China}

\author{Tommaso Morresi}
\email{morresi@ectstar.eu}
\affiliation{European Centre for Theoretical Studies in Nuclear Physics and Related Areas (ECT*),  Fondazione Bruno Kessler, Italy}

\author{Hongwei Xiong}
\email{xionghongwei@hbpu.edu.cn}
\affiliation{Center for Fundamental Physics, Hubei Polytechnic University, Huangshi 435003, China}
\date{\today}

\begin{abstract}
For strongly quantum-degenerate systems at finite temperatures, the fermion sign problem remains the major obstacle to first-principles simulations. In this work, we apply the recently proposed pseudo-fermion method—designed to overcome the sign problem—to strongly quantum-degenerate uniform electron gases. We find that the pseudo-fermion method can efficiently and highly accurately infer the energy of the uniform electron gas while being free from the fermion sign problem. For example, in the strongly quantum-degenerate regime where RPIMC fails (33 spin-polarized electrons at the density parameter $r_s=0.5$), the relative deviation between the pseudo-fermion method and the exact CPIMC result is only $0.6\%$. In particular, the pseudo-fermion method bridges the gap where neither CPIMC nor RPIMC can accurately simulate the regime $1 \leq r_s \leq 2$ at the reduced temperature $\theta = 0.0625$. This work demonstrates that the pseudo-fermion method opens a new pathway for studying strongly quantum-degenerate system in a sign-problem-free manner.
\end{abstract}

\maketitle

\section{Introduction}

When applying path-integral Monte Carlo (PIMC) \cite{CeperleyRMP,CeperleyBook,Tuckerman,Fosdick,Jordan,barker,Morita,Burov1,Burov2,Helium4} to quantum systems, the curse of dimensionality is overcome. However, for fermionic systems, the PIMC framework introduces the fermion sign problem \cite{Loh,ceperley,troyer,ZhangS, diagrammatic,Booth,Schoof,PB,PB2,PB3,Groth,Malone,Egger,Dornheim,Alex,Hou,WDM,HirshbergFermi,DornheimFermi} due to the antisymmetry of fermionic wavefunctions under exchange. For instance, first-principles simulations of uniform electron gases have long been severely hindered by the sign problem. When the Wigner–Seitz radius $r_s$ (density parameter) satisfies $0.1 \lesssim r_s \lesssim 10$, the system is considered to fall within the dense-matter regime \cite{WDM,Bonitz-Review,RoadMap}. Such dense matter plays an important role in inertial confinement fusion as well as in the interiors of many astrophysical objects \cite{RoadMap}. 

At finite temperatures, the restricted path integral Monte Carlo (RPIMC)\cite{RPIMC,Brown} is a popular method that avoids the sign problem during simulations; however, for the strongly quantum-degenerate uniform electron gas (UEG) with $r_s<2$, uncontrolled approximations pose obstacles to achieving high accuracy \cite{Bonitz-Review,CPIMC}. Although the configuration path integral Monte Carlo (CPIMC)\cite{Schoof,Groth,CPIMC} provides a highly accurate simulation method for the dense uniform electron gas, provided that the average sign remains manageable, it is still limited by the sign problem during simulations. For the case of $\theta = 0.0625$ and $r_s > 1$, the fermion sign problem in CPIMC becomes prohibitively severe, preventing reliable simulations for $33$ spin-polarized electrons, which constitutes a typical medium-scale UEG. Simultaneously, for $\theta = 0.0625$ and $r_s < 2$, the fixed-node approximation in RPIMC introduces notable systematic biases \cite{CPIMC}. Consequently, the regime defined by $\theta = 0.0625$ and $1 \leq r_s \leq 2$ represents a challenging gap where no existing method can currently provide reliable simulation results.
One of the objectives of this work is to fill the gap in the parameter regime that has so far been difficult for other methods to simulate.

In the context of sign-problem-free simulations of dense uniform electron gases, the main approaches developed so far include restricted path integral Monte Carlo (RPIMC) \cite{RPIMC,Brown} and fictitious identical particle methods \cite{XiongFSP}. The isothermal $\xi$-extrapolation scheme \cite{XiongFSP,Dornheim1} based on fictitious identical particles is suitable for weakly and moderately quantum-degenerate regimes but fails for strongly quantum-degenerate electron gases. For strongly quantum-degenerate electron gases, the constant energy $\xi$-extrapolation \cite{Xiong-xi,Morresi1,Morresi2,Yang} is required. RPIMC, on the other hand, breaks down for dense uniform electron gases at $r_s \lesssim 1$ due to its uncontrolled approximations \cite{Bonitz-Review}. Recently, pseudo-fermions \cite{pseudo}, introduced as a type of fictitious particles, have provided a new opportunity to overcome the fermion sign problem.

In this work, we perform numerical simulations of strongly quantum-degenerate uniform electron gases and demonstrate that the pseudo-fermion method \cite{pseudo} can efficiently and highly accurately reproduce the energy. For example, in the strongly quantum-degenerate case of $33$ spin-polarized electrons at $r_s = 0.5$, we find that the relative deviation between the pseudo-fermion method and exact result \cite{CPIMC} of CPIMC is only $0.6\%$. Regarding PIMC-based approaches for sign-problem-free simulations, this work shows that the pseudo-fermion method \cite{pseudo} provides a new alternative to both fictitious identical particles \cite{Dornheim1,Morresi2,Dornheim2,Dornheim3,Dornheim4,Dornheim5,Taylor,XiongGPU} and RPIMC \cite{Bonitz-Review,RoadMap,Brown,RPIMC} for uniform electron gases.

The remainder of this paper is organized as follows. In Sec. \ref{Pmethod}, we describe the general strategy of applying the pseudo-fermion method to uniform fermionic systems. In Sec. \ref{results}, we present energy calculations for typical cases of strongly quantum-degenerate uniform electron gases. Finally, Sec. \ref{summary} gives a brief summary and discussion.

\section{Application of the Pseudo-Fermion Method to the Uniform Electron Gas}
\label{Pmethod}

\subsection{A Heuristic Toy Model Leading to the Pseudo-Fermion Method}
\label{math}

We consider the following integral:
\begin{equation}
Z_F(\beta,a,\lambda)=\int_{-\infty}^{\infty}dx~xe^{-\beta (x-a)^2}(1+\lambda e^{-x^2}).
\label{ficP}
\end{equation}
The integrand above can be either positive or negative over the integration domain, and in the limit $a \rightarrow 0$, we have $Z_F \rightarrow 0$. We now attempt to compute
\begin{equation}
E_f(\beta,a,\lambda)=-\frac{\partial\ln Z_F(\beta,a,\lambda)}{\partial\beta}.
\end{equation}
Immediately, we obtain
\begin{equation}
E_f(\beta,a,\lambda)=\frac{\int_{-\infty}^{\infty}dx~ (x-a)^2xe^{-\beta (x-a)^2}(1+\lambda e^{-x^2})}{\int_{-\infty}^{\infty}dx~ xe^{-\beta (x-a)^2}(1+\lambda e^{-x^2})}.
\end{equation}
Our aim is to demonstrate how Monte Carlo importance sampling can be employed to evaluate $E_f(\beta, a, \lambda)$.

However, we immediately encounter a difficulty: the common factor in both numerator and denominator,
$x \, e^{-\beta (x-a)^2}\bigl(1+\lambda e^{-x^2}\bigr)$,
can take both positive and negative values, which makes it impossible to carry out importance sampling over the entire domain. To enable Monte Carlo importance sampling, we attempt the following approach. We rewrite $E_f(\beta,a,\lambda)$ as
\begin{equation}
E_f(\beta,a,\lambda)=\frac{A(\beta,a,\lambda)}{X(\beta,a,\lambda)}.
\label{direct}
\end{equation}
Here,
\begin{equation}
A(\beta,a,\lambda)=\frac{\int_{-\infty}^{\infty}dx~\frac{x}{|x|}(x-a)^2 |x|e^{-\beta (x-a)^2}(1+\lambda e^{-x^2})}{\int_{-\infty}^{\infty}dx~ |x|e^{-\beta (x-a)^2}(1+\lambda e^{-x^2})},
\end{equation}
\begin{equation}
X(\beta,a,\lambda)=\frac{\int_{-\infty}^{\infty}dx~\frac{x}{|x|} |x|e^{-\beta (x-a)^2}(1+\lambda e^{-x^2})}{\int_{-\infty}^{\infty}dx~ |x|e^{-\beta (x-a)^2}(1+\lambda e^{-x^2})}.
\label{signX}
\end{equation}
We refer to $X$ as the sign factor, since $\frac{x}{|x|}$ is the sign function. In the limit $a \rightarrow 0$, we have $X \rightarrow 0$.

For the expression of $E_f(\beta,a,\lambda)$ given in Eq.~(\ref{direct}), we can perform Monte Carlo importance sampling on the nonnegative function
$|x| e^{-\beta (x-a)^2}\bigl(1+\lambda e^{-x^2}\bigr)$
over the entire domain, thereby yielding Monte Carlo estimates $\bar{A}$ and $\bar{X}$ for $A(\beta,a,\lambda)$ and $X(\beta,a,\lambda)$, respectively. This yields an approximate value for $E_f(\beta,a,\lambda)$ as $\bar{A}(\beta,a,\lambda)/\bar{X}(\beta,a,\lambda)$.

Yet, we immediately face a serious problem: if $X(\beta,a,\lambda)$ is extremely small, then it is impossible to obtain accurate Monte Carlo estimates of $A(\beta,a,\lambda)$ and $X(\beta,a,\lambda)$, which may lead to a huge bias in the simulation of $E_f(\beta,a,\lambda)$. Consider, for instance, the case $\beta=1, a=10^{-10}, \lambda=0.1$. In this case, performing the integration without resorting to Monte Carlo methods shows that  $X = 1.74773\times 10^{-10}$. Let the number of Monte Carlo samples be $G$. Then the statistical fluctuation due to importance sampling scales as $\Delta X \sim 1/\sqrt{G}$. This implies that $G \gg 10^{20}$ is required to meet accuracy requirements. Such a seemingly simple Monte Carlo problem is already beyond the capabilities of current supercomputers.

We now consider the case where $E_f(\beta,a,\lambda=0)$ is known in advance. Then a natural question arises: can we still obtain highly accurate Monte Carlo estimates of $E_f(\beta,a,\lambda)$ when $\lambda \neq 0$?

By inspecting Eq.~(\ref{ficP}) for $Z_F(\beta,a,\lambda)$, we construct the following expression by taking the absolute value of the integrand:
\begin{equation}
Z_{pf}(\beta,a,\lambda)=\int_{-\infty}^{\infty}dx~|x|e^{-\beta (x-a)^2}(1+\lambda e^{-x^2}).
\end{equation}
We then define
\begin{equation}
E_{pf}(\beta,a,\lambda)=-\frac{\partial\ln Z_{pf}(\beta,a,\lambda)}{\partial\beta}.
\end{equation}
The quantity $E_{pf}(\beta,a,\lambda)$ can be evaluated exactly by Monte Carlo sampling.

We write
\begin{equation}
Z_F(\beta,a,\lambda)=X(\beta,a,\lambda)Z_{pf}(\beta,a,\lambda).
\end{equation}
In particular, for $\lambda=0$,
\begin{equation}
Z_F(\beta,a,\lambda=0)=X(\beta,a,\lambda=0)Z_{pf}(\beta,a,\lambda=0).
\end{equation}
Since the integrand of $Z_F$ can be positive or negative, it is easy to encounter cases with $|X| \ll 1$. For example, when $\beta=1$ and $a=10^{-10}$, performing the integration without resorting to Monte Carlo methods yields $X(\beta,a,\lambda=0)=1.77245\times 10^{-10}$.

We further set
\begin{equation}
Z_F(\beta,a,\lambda)=X(\beta,a,\lambda=0)\mathcal{O}(\beta,a,\lambda)Z_{pf}(\beta,a,\lambda).
\end{equation}
Here, $\mathcal{O}(\beta,a,\lambda)$ denotes the correction to $X(\beta,a,\lambda=0)$. Since the sign of the integrand in $Z_F(\beta,a,\lambda)$ originates from the factor $x e^{-\beta (x-a)^2}$, we expect the factor $(1+\lambda e^{-x^2})$ to act only as a correction in some cases. Noting that $\mathcal{O}(\beta,a,\lambda=0)=1$, we anticipate situations where $\mathcal{O}(\beta,a,\lambda)\sim O(1)$. That is,
\begin{equation}
|\ln X(\beta,a,\lambda=0)|\gg|\ln \mathcal{O}(\beta,a,\lambda)|.
\label{appO}
\end{equation}

We now define
\begin{equation}
E_X(\beta,a,\lambda=0)=-\frac{\partial\ln X(\beta,a,\lambda=0)}{\partial\beta},
\end{equation}
\begin{equation}
E_\mathcal{O}(\beta,a,\lambda)=-\frac{\partial\ln \mathcal{O}(\beta,a,\lambda)}{\partial\beta},
\end{equation}
\begin{equation}
E_{pf}(\beta,a,\lambda)=-\frac{\partial\ln Z_{pf}(\beta,a,\lambda)}{\partial\beta}.
\end{equation}
Thus
\begin{equation}
E_f(\beta,a,\lambda)=E_X(\beta,a,\lambda=0)+E_\mathcal{O}(\beta,a,\lambda)+E_{pf}(\beta,a,\lambda).
\end{equation}

When the condition Eq.~(\ref{appO}) is satisfied, we may approximate
\begin{equation}
E_f(\beta,a,\lambda)\approx E_X(\beta,a,\lambda=0)+E_{pf}(\beta,a,\lambda).
\end{equation}
Using
\begin{equation}
E_f(\beta,a,\lambda=0)= E_X(\beta,a,\lambda=0)+E_{pf}(\beta,a,\lambda=0),
\end{equation}
we define
\begin{equation}
\tilde E_f(\beta,a,\lambda)= E_f(\beta,a,\lambda=0)+(E_{pf}(\beta,a,\lambda)-E_{pf}(\beta,a,\lambda=0)),
\label{tildeE}
\end{equation}
which serves as an approximation of $E_f(\beta,a,\lambda)$. The quantity $\tilde E_f(\beta,a,\lambda)$ represents the estimate of $E_f$ inferred by the present method. As noted earlier, $E_f(\beta,a,\lambda=0)$ is assumed to be known a \textit{priori}, while the correction term ($E_{pf}(\beta,a,\lambda)-E_{pf}(\beta,a,\lambda=0)$) can be exactly evaluated via Monte Carlo sampling. In this approximation, we neglect the contribution of $E_\mathcal{O}(\beta,a,\lambda)$.

\begin{figure}[htbp]
\begin{center}
\includegraphics[width=0.8\textwidth]{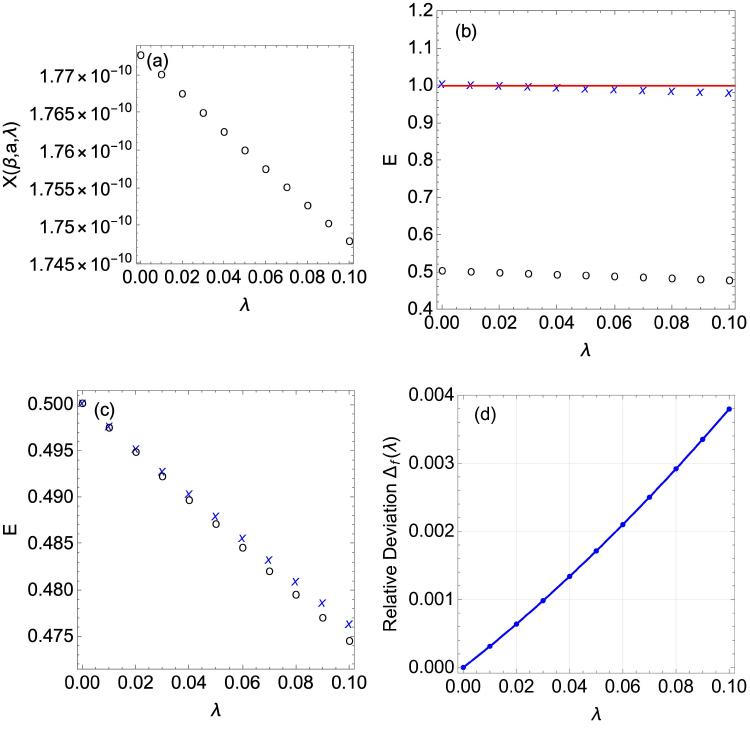}
\caption{\label{ToyModel} For $\beta=1,a=10^{-10}$: (a) the sign factor $X(\beta,a,\lambda)$; (b) blue crosses: $E_{pf}(\lambda)$, red line: $E_{pf}(\lambda=0)$, black circles: $E_f(\lambda)$; (c) blue crosses: $\tilde E_f$, black circles: $E_f$; (d) relative deviation between $\tilde E_f$ and $E_f$.}
\end{center}
\end{figure}

To illustrate the method, we take $\beta=1$ and $a=10^{-10}$ as an example. In Fig.\ref{ToyModel}(a), the black circles show the magnitude of the sign factor $X(\beta,a,\lambda)$ for different $\lambda$. We emphasize again that, for such extremely small sign factors, it is impossible to accurately infer $E_f$ via Monte Carlo sampling based on Eq.(\ref{signX}). In Fig.\ref{ToyModel}(b), the black circles denote exact values of $E_f(\beta,a,\lambda)$ (obtained without Monte Carlo), while the blue crosses and red line represent $E_{pf}(\beta,a,\lambda)$ for different $\lambda$, and $E_{pf}(\beta,a,\lambda=0)$, respectively. We observe that $E_{pf}(\beta,a,\lambda)$ deviates significantly from the desired $E_f(\beta,a,\lambda)$, though both display a similar monotonic decrease with increasing $\lambda$. In Fig.\ref{ToyModel}(c), the blue crosses represent $\tilde E_f(\beta,a,\lambda)$, while the black circles represent $E_f(\beta,a,\lambda)$. The two are seen to be very close. Finally, in Fig.~\ref{ToyModel}(d), we show the relative deviation
\begin{equation}
\Delta_f(\lambda)=\frac{\tilde E_{f}(\beta,a,\lambda)-E_f(\beta,a,\lambda)}{E_f(\beta,a,\lambda)}.
\end{equation}
We note that the relative deviation gradually increases with $\lambda$, but remains below $0.4\%$ at $\lambda=0.1$.

Of course, for this one-dimensional integral problem, Monte Carlo methods are unnecessary. For convenience, none of the results in Fig.~\ref{ToyModel} were obtained \textit{via} Monte Carlo. However, for higher-dimensional or multidimensional problems, there are generally no reliable alternatives to Monte Carlo. For example, in the path-integral representation of many-body quantum system, its evaluation must rely on Monte Carlo methods. The strategy introduced here for one-dimensional integrals can be extended to higher dimensions, thus providing a new avenue for overcoming the fermion sign problem.

The deliberately chosen symbols here are meant to establish a connection with fermionic systems. In Table~\ref{corres}, we summarize the correspondence. From the path-integral representation of the fermionic partition function, it is straightforward to establish the link between this simple example and fermions, and from there to develop various pseudo-fermion-based approaches for inferring fermionic energies. In this paper, we propose one such pseudo-fermion method for uniform fermionic systems, while not excluding the possibility that more advanced pseudo-fermion approaches may be developed in the future.

\begin{table}[h]
    \centering
    \begin{tabular}{|c|c|c|}
        \hline
        Symbol & Fermi system\\
        \hline
        $a$ & Parameter of noninteracting fermions\\
        $\lambda$ & Coupling constant between particles\\
        $\beta$ & $\beta=1/k_BT$ ($k_B\rightarrow$Boltzmann constant, $T\rightarrow$temperature)\\
        $Z_{F}(\beta,a,\lambda)$ & Partition function of interacting fermions\\
         $Z_{F}(\beta,a,\lambda=0)$& Partition function of noninteracting fermions\\
         $E_F=-\frac{\partial\ln Z_{F}}{\partial \beta}$& Energy of fermions\\
        $Z_{pf}(\beta,a,\lambda)$ & Partition function of pseudo-fermions\\
        $E_{pf}=-\frac{\partial\ln Z_{pf}}{\partial\beta}$& Energy of pseudo-fermions\\
        $X(\beta,a,\lambda)$&Sign factor of fermions\\
        $X(\beta,a,\lambda=0)$&Sign factor of noninteracting fermions\\
        Eq. (\ref{direct})& Direct PIMC simulation of fermions\\
        $\mathcal{O}(\beta,a,\lambda)$& The correction to sign factor of noninteracting fermions\\
        $\tilde {E}_f(\beta,a,\lambda)$ & The inferred fermion energy based on pseudo-fermions\\
        \hline
    \end{tabular}
    \caption{The connection between the simple one-dimensional example and the pseudo-fermion method for fermionic systems.}
    \label{corres}
\end{table}

\subsection{Model of the Polarized Uniform Electron Gas}

The analysis in this work generally applies to fermions in a periodic box. Here, we consider a spin-polarized uniform electron gas in a periodic box with Coulomb repulsion. We adopt atomic units and the unit of energy is the Hartree energy in this work. In addition, we adopt the convention $k_B = 1$. For the uniform electron gas, the Wigner–Seitz radius $r_s$ (density parameter) is defined by $4\pi r_s^3/3=1/n$, where $n$ is the density of the uniform electron gas \cite{WDM}. For the spin-polarized uniform electron gas, the Fermi energy is $E_F=q_F^2/2$, with $q_F=(6\pi^2 n)^{1/3}$. We further define the dimensionless temperature $\theta=T/E_F$.

The Hamiltonian defining uniform electron gas with periodic boundary condition is given by \cite{WDM}
\begin{equation}
\hat H=-\frac{1}{2}\sum_{i=1}^N\nabla_i^2+\hat V_{\text{int}}.
\end{equation}
Here
\begin{equation}
\hat V_{\text{int}}=\sum_{k>j}^NW_E(\mathbf{r}_k,\mathbf{r}_j)+\frac{N}{2}\xi_M,
\end{equation}
where the Ewald potential is defined as
\begin{equation}
W_E(\mathbf{r}_k,\mathbf{r}_j)=\frac{1}{V\pi}\sum_{\mathbf{G}\neq0}(G^{-2}e^{-\frac{\pi^2G^2}{\kappa^2}+2\pi i\mathbf{G}\cdot(\mathbf{r}_k-\mathbf{r}_j)})-\frac{\pi}{\kappa^2V}+\sum_{\mathbf{R}\neq0}\frac{\text{erfc}(\kappa|\mathbf{r}_k-\mathbf{r}_j+\mathbf{R}|)}{|\mathbf{r}_k-\mathbf{r}_j+\mathbf{R}|},
\end{equation}
where $\mathbf{G}=\textbf{n}L^{-1}$ and $\mathbf{R}=\textbf{m}L$ for integer vectors $\textbf{n}$ and $\textbf{m}$, $\kappa$ is an optimization parameter that can be set so that the above summations converge in a fast manner. Moreover, the Madelung constant $\xi_M$ is defined by
\begin{equation}
\xi_M=\lim_{\mathbf{r}\rightarrow\mathbf{s}}(W_E(\mathbf{r},\mathbf{s})-\frac{1}{|\mathbf{r}-\mathbf{s}|}).
\end{equation}
For simplicity and convenience, we consider here the spin-polarized uniform electron gas, since extending the pseudo-fermion method to finite spin polarization in a spin-independent Hamiltonian is straightforward.

\subsection{Partition Function of Pseudo-Fermions in a Periodic Box}

The partition function of fermions is given by
\begin{equation}
Z_F(\beta, \lambda) = \mathrm{Tr}\left( e^{-\beta \hat{H}} \right).
\end{equation}
Here, $ \beta = 1 / ({k_B} T) $, where $ T $ is the temperature and $ k_B $ is the Boltzmann constant ($k_B$ has already been set equal to 1 above). 
$\lambda$ represents the coupling strength between particles, with $\lambda=0$ corresponding to an ideal Fermi gas. Below, we present the approach to deriving the fermionic partition function within the path-integral framework, and in Appendix \ref{appendix}, we provide some additional derivation details.

We can write the partition function of the fermionic system with periodic boundary condition as:
\begin{equation}
Z_F(\beta, \lambda) =\int d\textbf{R} \sum_P\sum_{\textbf{W}} (-1)^{N_P} \left< P\textbf{R},\textbf{W} \right| e^{-\beta \hat{H}} \left| \textbf{R},0 \right>,
\end{equation}
where $\textbf{W}=(\mathbf{w}_1,...,\mathbf{w}_N)$ is a vector of integers which can take any value in ($-\infty$,$\infty$). Here, \( \textbf{R}\equiv (\textbf{r}_1, \cdots, \textbf{r}_N)\) includes the coordinates of all \( N \) particles, and $\left| \textbf{R},\textbf{W} \right>$ represents the position vector $(\textbf{r}_1+\mathbf{w}_1L, \cdots, \textbf{r}_N+\mathbf{w}_NL)$ for $L$ the box length. \( P \) represents the permutation operator acting on the coordinates, and \( N_P \) denotes the minimal number of pairwise exchanges required to restore the original order of the coordinates under permutation \( P \). 
The presence of the factor \( (-1)^{N_P} \) in the fermionic partition function \( Z_F(\beta, \lambda) \) leads to the fermion sign problem \cite{WDM,Dornheim}, compared with the partition function of bosons.

We can express the partition function of the fermionic system as:
\begin{equation}
Z_F(\beta, \lambda) = \int d\textbf{R} \sum_P\sum_{\textbf{W}} (-1)^{N_P} \left< P\textbf{R},\textbf{W} \right| e^{-\Delta\tau \hat{H}} \cdots e^{-\Delta\tau \hat{H}} \left| \textbf{R},0 \right>.
\label{Fdelta}
\end{equation}
Here, \( \Delta \tau = \beta / M \).
When \( \Delta \tau \) is small, we can apply the Suzuki-Trotter decomposition \cite{Suzuki} and insert appropriate identity operators over momentum and position to transform the above partition function into a high-dimensional integral.

For the fermionic partition function above, we have two ways to express it:

(1) By inserting the following identity operator for distinguishable particles
\begin{equation}
\hat I_D=\int d\textbf {R}\sum_{{\textbf W}_1,{\textbf W}_2}\left |\textbf R, {\textbf W}_1\right>\left<\textbf R,{\textbf W}_2\right |,
\end{equation}
we can obtain the conventional expression \cite{Dornheim} for the fermionic partition function . 

The use of the identity operator for distinguishable particles, $\hat{I}_D$, is a valid mathematical convenience that respects the principle of indistinguishability. This is justified because the set of distinguishable states $\{|\mathbf{R}, \mathbf{W}\rangle\}$ forms a complete basis for the full Hilbert space, which contains the antisymmetric fermionic subspace as a subset. By inserting $\hat{I}_D$ as a resolution of identity, we project the propagator onto a basis of definite coordinates, effectively decomposing the path integral into discrete spatial configurations. The physical requirement of indistinguishability is then rigorously recovered \textit{via} the antisymmetrizer $\hat{A} = \sum_P (-1)^{N_P} \hat{P}$ when evaluating the partition function $Z_F = \mathrm{Tr}(\hat{A} e^{-\beta \hat{H}})$. Given that $\hat{H}$ exhibits permutation symmetry, this procedure is mathematically equivalent to summing over all permutations of particle labels with the appropriate sign, effectively filtering the fermionic contribution from the total sum of paths.

In this fermionic partition function, the factor $\sum_P (-1)^{N_P}$ acts only on the beads in one of the imaginary time slices. For this case, if we replace the factor $\sum_P (-1)^{N_P}$ with $\sum_P$ we then obtain the bosonic partition function $Z_B(\beta,\lambda,M)$.
We define \cite{Dornheim}
\begin{equation}
Z_F(\beta,\lambda,M)=S(\beta,\lambda,M)Z_B(\beta,\lambda,M).
\label{Sfactor}
\end{equation}
Here, $S(\beta, \lambda, M)$ is the so-called sign factor.

(2) Although the above approach to handling the fermionic partition function is popular \cite{Dornheim}, there exists an alternative formulation that, from a formal standpoint, preserves more of the exchange antisymmetry of fermions and can be regarded as more elegant in a certain sense.

Fully taking into account the exchange antisymmetry of fermions, we can define the coordinate basis as follows:
\begin{equation}
\left |\textbf R\right >_F=\frac{1}{\sqrt N!}\sum_P (-1)^{N_P}\sum_{\textbf W}\left|P \textbf R,\textbf W\right>.
\label{Rcoordinate}
\end{equation}
For this coordinate basis, due to the factor $(-1)^{N_P}$, if any two fermions occupy the same position, the corresponding coordinate basis state does not exist. 

The vanishing of the state when two fermions occupy the same position is a direct consequence of the exchange antisymmetry encoded in Eq. (\ref{Rcoordinate}). Specifically, if two fermions $i$ and $j$ are at the same position ($\mathbf{r}_i = \mathbf{r}_j$), then for every permutation $P$ in the sum, there exists a corresponding permutation $P' = P \circ \tau_{ij}$ (where $\tau_{ij}$ is the transposition of particles $i$ and $j$). Because $\mathbf{r}_i = \mathbf{r}_j$, the spatial configuration remains identical: $|P\mathbf{R}, \mathbf{W}\rangle = |P'\mathbf{R}, \mathbf{W}\rangle$. However, their parity is opposite, \textit{i.e.}, $(-1)^{N_{P'}} = -(-1)^{N_{P}}$.

Consequently, these two terms cancel each other out:
\begin{equation}
(-1)^{N_P} |P\mathbf{R}, \mathbf{W}\rangle + (-1)^{N_{P'}} |P'\mathbf{R}, \mathbf{W}\rangle = 0.
\end{equation}
Since the entire sum over $N!$ permutations can be partitioned into such canceling pairs, the total state $|\mathbf{R}\rangle_F$ becomes zero. This is the mathematical representation of the Pauli exclusion principle in the coordinate basis.

From this coordinate basis, we obtain the following form of the identity operator:
\begin{equation}
\hat I_F=\int d\textbf R \left |\textbf R\right >_F \left <\textbf R\right |_F.
\label{Fidentity}
\end{equation}
By inserting the above identity operator into Eq. (\ref{Fdelta}), we can obtain an alternative integral representation of the fermionic partition function.

For the case of a periodic box, we have
\begin{equation}
\begin{split}
\sum_P\sum_{\textbf{W}} (-1)^{N_P}\left< P\textbf{R},\textbf{W} \right|e^{-\Delta\tau \hat{H}}\left| \textbf{R}',0 \right> {\simeq} &\frac{1}{(2\pi\Delta\tau)^{Nd/2}}\sum_P\sum_{\textbf{W}} (-1)^{N_P}e^{-\frac{1}{2\Delta\tau}\left(P\textbf {R}-\textbf {R}'+\textbf{W}L\right)^2}\\&e^{-\frac{\Delta\tau}{2}(V_{\text{int}}(\mathbf{R}+\mathbf{W}L)+V_{\text{int}}(\mathbf{R}'))}+O(\Delta\tau^3).
\end{split}
\label{RR}
\end{equation}
For periodic systems the interaction potential satisfies $V_{\text{int}}(\mathbf{R}+\mathbf{W}L)=V_{\text{int}}(\mathbf{R})$, so there is no summation over $\mathbf{W}$ for the potential part. When Dornheim \textit{et al.} \cite{DornheimPB} applied permutation blocking path integral Monte Carlo (PB-PIMC) to the uniform electron gas, they treated the interactions between electrons in the periodic box in the same way.

Based on Eqs. (\ref{Fdelta}), (\ref{Fidentity}), and (\ref{RR}), we have the following approximate fermionic partition function for finite $M$
\begin{equation}
Z_F(\beta, \lambda,M) = \int \prod_{j=1}^{M} d\textbf R^j D_{\mathrm{free}}(\textbf{R}^j, \textbf{R}^{j+1}; \Delta\tau)
 e^{-\Delta\tau V_{\mathrm{int}}(\textbf{R}^j,\lambda)}.
\label{fermionZ}
\end{equation}
Here \( \textbf{R}^j \) denotes the coordinates of all the beads in the \( j \)-th imaginary time slice, where \( j = 1, 2, \cdots, M \). In this article, we adopt the convention $\textbf R^1\equiv \textbf R^{M+1}$. $V_{\mathrm{int}}(\textbf{R}^j,\lambda)$ is the interaction potential between particles with $\lambda$ being the coupling constant, for Coulomb interaction $\frac{\lambda}{|\textbf{r}_i-\textbf{r}_j|}$ studied in this work, we have $\lambda=1$. To formulate the pseudo-fermion method, we treat $\lambda$ as a continuously adjustable real coupling constant and assume a general interaction potential of the form $V_{\mathrm{int}}(\mathbf{R}^j, \lambda) = \lambda V_{\mathrm{int}}(\mathbf{R}^j)$.

In Eq. (\ref{fermionZ}), the antisymmetric free-fermion propagator \cite{latticeFermi,Takahashi,Miura,Lyubartsev1,Lyubartsev2,Imada,Filinov1,Filinov2,ChinF} is:
\begin{equation}
D_{\mathrm{free}}(\textbf{R}^j, \textbf{R}^{j+1}; \Delta\tau) = \textit{det}\left(\frac{1}{(2\pi\Delta\tau)^{d/2}}\sum_{\textbf{W}}\exp\left(-\frac{1}{2\Delta\tau}\left(\textbf {R}^j_{l}-\textbf {R}^{j+1}_m+\textbf{W}L\right)^2\right)\right).
\label{freeD}
\end{equation}
Here $\textit{det}$ represents determinant (in the above expression, the indices of the matrix are $l,m$), while $d$ denotes the spatial dimension of the fermionic system. In $\textbf{R}^j_l$, the superscript $j$ indicates the $j$-th imaginary time slice, and the subscript $l$ labels the $l$-th particle. We sum over all integer vectors $\textbf{W}$ where each component is between $-w_{\text{max}}$ and $w_{\text{max}}$ for some cutoff $w_{\text{max}}$.
For the two- and three-dimensional cases \cite{Chin1,Chin2}, for different $\textbf{R}^j$ and $\textbf{R}^{j+1}$, $D_{\text{free}}(\textbf{R}^j, \textbf{R}^{j+1}; \Delta\tau)$ can be either positive or negative, which prevents us from performing importance sampling of the high-dimensional integral in Eq. (\ref{fermionZ}). Compared with inserting the identity operator $\hat I_D$, the fermionic partition function given by Eq. (\ref{fermionZ}) is formally more elegant, since each imaginary-time slice is treated in the same manner. As $M\rightarrow\infty$, we formally arrive at the exact fermionic partition function $Z_F(\beta,\lambda)$.

In the second form of the fermionic partition function considered here, the sign problem still persists.
In order to completely avoid the sampling difficulties caused by the sign problem during numerical simulations, we define the pseudo-fermion partition function \cite{pseudo} as follows:
\begin{equation}
Z_{pf}(\beta, \lambda,M) = \int \prod_{j=1}^{M}d\textbf R^j \left |D_{\mathrm{free}}(\textbf{R}^j, \textbf{R}^{j+1}; \Delta\tau)\right |
 e^{-\Delta\tau V_{\mathrm{int}}(\textbf{R}^j,\lambda)}.
\label{pseudofermion}
\end{equation}
The integrand in the pseudo-fermion partition function is always positive, allowing for importance sampling using Monte Carlo methods. For the pseudo-fermion partition function, we note that within each imaginary-time slice, no two beads can occupy the same position, meaning that a Pauli-like exclusion principle similar to that of fermions is still satisfied. For this reason, we refer to the fictitious particles constructed here as pseudo-fermions.

The case of $M=2$ deserves special consideration. For fermions, we have:
\begin{equation}
Z_F(\beta, \lambda, M=2) = \int d\textbf R^1 d\textbf R^2 D_{\mathrm{free}}(\mathbf{R}^1, \mathbf{R}^{2}; \Delta\tau)
D_{\mathrm{free}}(\mathbf{R}^2, \mathbf{R}^{1}; \Delta\tau)
e^{-\Delta\tau (V_{\mathrm{int}}(\mathbf{R}^1,\lambda)+V_{\mathrm{int}}(\mathbf{R}^2,\lambda))/2}.
\label{fermionZ2}
\end{equation}
Since $D_{\mathrm{free}}(\mathbf{R}^1, \mathbf{R}^{2}; \Delta\tau) = D_{\mathrm{free}}(\mathbf{R}^2, \mathbf{R}^{1}; \Delta\tau)$, we note that the Monte Carlo sampling of $Z_F(\beta, \lambda, M=2)$ does not suffer from the sign problem. Unfortunately, in the presence of interactions, the above partition function is only a rough approximation to the true fermionic partition function (the Suzuki-Trotter decomposition \cite{Suzuki} requires sufficiently large $M$ to be valid). Once $M$ is increased to handle interactions more accurately, the sign problem emerges in two and three dimensions. In the absence of interactions ($\lambda=0$), however, the above expression is exact, allowing the exact simulation of the thermodynamic properties of noninteracting fermions in the canonical ensemble. The goal of the pseudo-fermion method is to enable the use of larger $M$ values while avoiding the sign problem during simulations.

\subsection{General Strategy of the Pseudo-Fermion Method for the Uniform Electron Gas}
\label{Mcana}

For the uniform electron gas, we set
\begin{equation}
Z_F(\beta,\lambda) = X(\beta,\lambda,M) Z_{pf}(\beta,\lambda,M).
\label{Fpartition}
\end{equation}
Here, $Z_F(\beta,\lambda)$ denotes the exact fermionic partition function, which is independent of $M$. $X(\beta,\lambda,M)$ is the sign factor comparing the pseudo-fermion and fermion partition functions. This factor also incorporates the approximation arising from the Suzuki-Trotter decomposition \cite{Suzuki} at finite $M$. For the noninteracting case, we define
\begin{equation}
Z_F(\beta,\lambda=0) = X(\beta,\lambda=0,M) Z_{pf}(\beta,\lambda=0,M).
\end{equation}
Here, $X(\beta,\lambda=0,M)$ is the sign factor without interparticle interactions.

We rewrite Eq.(\ref{Fpartition}) as
\begin{equation}
Z_F(\beta,\lambda) = X(\beta,\lambda=0,M) \mathcal{O}(\beta,\lambda,M) Z_{pf}(\beta,\lambda,M).
\label{ZfO}
\end{equation}
Here, $\mathcal{O}(\beta,\lambda,M)$ represents the correction to the sign factor due to interparticle interactions. For $\lambda=0$, $\mathcal{O}(\beta,\lambda=0,M)=1$. Since the signs in the path-integral expression of $Z_F(\beta,\lambda)$ are determined by the free-fermion propagator between neighboring imaginary-time slices (which is a sum of $N!$ permutation terms), and the interaction only  
modifies the magnitude of the weights rather than their signs, we expect $\mathcal{O}(\beta,\lambda,M)$ to be only a small correction to the sign factor, \textit{i.e.},
\begin{equation}
\left|\ln \mathcal{O}(\beta,\lambda,M)\right| \ll \left|\ln X(\beta,\lambda=0,M)\right|.
\label{smallO}
\end{equation}
Below, we will analyze the optimization of $M$ to minimize the contribution of $\mathcal{O}(\beta,\lambda,M)$ to the total energy.

The fermionic and pseudo-fermionic energies are defined as:
\begin{equation}
E_f(\beta,\lambda) = -\frac{\partial \ln Z_F(\beta,\lambda)}{\partial \beta},
\end{equation}
\begin{equation}
E_{pf}(\beta,\lambda,M) = -\frac{\partial \ln Z_{pf}(\beta,\lambda,M)}{\partial \beta}.
\end{equation}

For the noninteracting case ($\lambda=0$), at $M=2$, $Z_{pf}(\beta,\lambda=0,M=2)$ coincides with the exact fermionic partition function. At $M=2$, no sign problem exists, thus providing an exact method to simulate the energy of the noninteracting uniform electron gas. In this case, $X(\beta,\lambda=0,M=2)=1$, leading to
\begin{equation}
E_f(\beta,\lambda=0) = -\frac{\partial \ln Z_{pf}(\beta,\lambda=0,M=2)}{\partial \beta}.
\end{equation}

We further define
\begin{equation}
E_\mathcal{O}(\beta,\lambda,M) = -\frac{\partial \ln \mathcal{O}(\beta,\lambda,M)}{\partial \beta}.
\end{equation}
Hence,
\begin{equation}
E_f(\beta,\lambda) - E_f(\beta,\lambda=0) = \delta E_\mathcal{O}(\beta,\lambda,M) + \delta E_{pf}(\beta,\lambda,M),
\end{equation}
where
\begin{equation}
\delta E_\mathcal{O}(\beta,\lambda,M) = E_\mathcal{O}(\beta,\lambda,M) - E_\mathcal{O}(\beta,\lambda=0,M),
\end{equation}
\begin{equation}
\delta E_{pf}(\beta,\lambda,M) = E_{pf}(\beta,\lambda,M) - E_{pf}(\beta,\lambda=0,M).
\end{equation}
By the way, here $E_\mathcal{O}(\beta,\lambda=0,M)=0$ because $\mathcal{O}(\beta,\lambda=0,M)=1$.

Since both $E_f(\beta,\lambda)$ and $E_f(\beta,\lambda=0)$ are independent of $M$, we artificially treat $M$ as a continuous variable and obtain
\begin{equation}
\frac{\partial}{\partial M}\big(\delta E_\mathcal{O}(\beta,\lambda,M)+\delta E_{pf}(\beta,\lambda,M)\big)=0.
\end{equation}
In this case, we have
\begin{equation}
\left|\frac{\partial }{\partial M}\delta E_{pf}(\beta,\lambda,M)\right|=\left|\frac{\partial }{\partial M}\delta E_{\mathcal{O}}(\beta,\lambda,M)\right|.
\label{partialE}
\end{equation}

Since the simulation of $\delta E_{pf}(\beta, \lambda, M)$ is free from the sign problem, we can simulate $\delta E_{pf}(\beta, \lambda, M)$ efficiently and with high accuracy. In this work, the strategy we adopt is to simulate $\delta E_{pf}(\beta, \lambda, M)$ for various values of $M$ and search for the flat region, \textit{i.e.}, the minimum value of $\left|\frac{\partial }{\partial M}\delta E_{pf}(\beta, \lambda, M)\right|$. Given the relationship in Eq. ($\ref{partialE}$), it is expected that $\delta E_{\mathcal{O}}(\beta, \lambda, M)$ will exhibit a flat dependence on $M$, coinciding with the flat region of $\delta E_{pf}(\beta, \lambda, M)$. Since the change in the $M$ direction is slowest in this flat region, we reasonably hypothesize that the change in $\delta E_{\mathcal{O}}(\beta, \lambda, M)$ with respect to the $\lambda$ direction is also slow in this region. Considering that $\delta E_{\mathcal{O}}(\beta, \lambda=0, M)=0$, we approximate $\delta E_{\mathcal{O}}(\beta,\lambda\neq 0,M)$ to be zero within this flat region. In other words, we hope that for an appropriate value of $M$, $\mathcal{O}(\beta, \lambda, M)$ will only weakly depend on $\lambda$, such that this term's correction to the energy can be neglected.
In this case, the fermionic energy can be approximated by
\begin{equation}
E_f(\beta,\lambda)\approx E_f(\beta,\lambda=0)+\delta E_{pf}(\beta,\lambda;\text{flat}).
\label{approximate}
\end{equation}
Here, $\delta E_{pf}(\beta,\lambda;\text{flat})$ denotes $\delta E_{pf}$ in the flat region.
In the subsequent numerical simulations, we find that $\delta E_{pf}(\beta,\lambda,M)$, starting from $M=2$, indeed enters a plateau (flat) region as $M$ increases. Since fluctuations are always present in the simulations, we find in Sec. \ref{results} that making use of this plateau region, rather than relying on a single specific $M$ in the plateau region, is more advantageous for inferring the fermionic energy. 

In the plateau region, we expect $E_{\mathcal{O}}(\beta,\lambda;\text{flat})\equiv \delta E_{\mathcal{O}}(\beta,\lambda;\text{flat})$ to be minimized and thus negligible.
The relative deviation introduced by the pseudo-fermion method is
\begin{equation}
\Delta_f(\beta,\lambda;\text{flat}) = \frac{\delta E_\mathcal{O}(\beta,\lambda;\text{flat})}{E_f(\beta,\lambda)} = \frac{\delta E_\mathcal{O}(\beta,\lambda;\text{flat})}{\delta E_{pf}(\beta,\lambda;\text{flat})} \cdot \frac{\delta E_{pf}(\beta,\lambda;\text{flat})}{E_f(\beta,\lambda)}.
\label{deltaf}
\end{equation}
As noted earlier, $\delta E_\mathcal{O}(\beta,\lambda;\text{flat})$ arises from interaction-induced corrections to the sign factor, while $\delta E_{pf}(\beta,\lambda;\text{flat})$ corresponds to the energy difference between interacting and noninteracting pseudo-fermions. Therefore, it is reasonable to assume
\begin{equation}
\left|\frac{\delta E_\mathcal{O}(\beta,\lambda;\text{flat})}{\delta E_{pf}(\beta,\lambda;\text{flat})}\right| \ll 1.
\end{equation}
Of course, approximating the fermionic energy by Eq. (\ref{approximate}) involves two approximations: (1) neglecting the deviation due to $\delta E_\mathcal{O}(\beta,\lambda;\text{flat})$; and (2) the fact that the Suzuki-Trotter decomposition \cite{Suzuki} is only exact in the limit $M \to \infty$, while finite $M$ always introduces some bias.

Let us now revisit Eq.~(\ref{ZfO}).
$Z_F(\beta,\lambda)$ is the exact fermionic partition function.
We define Eq.~(\ref{pseudofermion}) as the pseudo-fermion partition function for a given $M$, without taking into account the approximation introduced by the Suzuki–Trotter decomposition. In this case, $\mathcal{O}(\beta,\lambda,M)$ serves both as a correction to the sign factor $X(\beta,\lambda=0,M)$ and as an inclusion of the approximation arising from the Suzuki–Trotter decomposition.
In the pseudo-fermion method, what we neglect is precisely the contribution $\delta E_\mathcal{O}(\beta,\lambda,M)$ that originates from $\mathcal{O}(\beta,\lambda,M)$. Therefore, when choosing the plateau region, we are in fact simultaneously aiming to minimize the approximation due to the Suzuki–Trotter decomposition.

At the current stage, neglecting $\delta E_\mathcal{O}(\beta,\lambda,M)$ in the plateau region of $\delta {E}_f(\beta,\lambda,M)$ is, from a strict mathematical perspective, merely a seemingly reasonable conjecture. 
Even though we have provided both mathematical and physical arguments for the applicability of the pseudo-fermion method, the method remains in a developmental stage, and consequently, its reliability and applicability must be carefully scrutinized.
Similar to the isothermal $\xi$-extrapolation method based on fictitious identical particles applied to the uniform electron gas by Dornheim \textit{et al.} \cite{Dornheim1,Dornheim2,Dornheim3,Dornheim4,Dornheim5,Taylor}, the strategy here is as follows: apply the pseudo-fermion method to small fermionic systems, compare with more expensive exact results to assess the deviation, and, once validated for small systems, extend the pseudo-fermion method to larger systems. Since the pseudo-fermion method avoids the sign problem during simulations, it enables the efficient treatment of large fermionic systems. In the next section (Sec.~\ref{results}), we further elaborate on the details of extracting fermionic energies from the pseudo-fermion method.

\subsection{Simulation techniques}
In order to perform Monte Carlo simulation for pseudo-fermions, we use standard Metropolis procedure with Markov chain to do the sampling. More details on the Monte Carlo moves we employed in the simulation and the formulas for their acceptance probability can be found in Ref \cite{pseudo}. Here, we present additional considerations specific to periodic boundary conditions.

To calculate the free-fermion propagator given by Eq. (\ref{freeD}), we need to evaluate each element of the matrix in its determinant, which for systems with periodic boundary condition is given by
\begin{equation}
A(\textbf {R}^j,\textbf {R}^{j+1})_{l,m}=\sum_{\mathbf{W}}\exp\left(-\frac{1}{2\Delta\tau}\left(\textbf {R}_{l}^j-\textbf {R}^{j+1}_m+\textbf{W}L\right)^2\right).
\label{Wsum}
\end{equation}
Here, $\textbf{R}^j$ and $\textbf{R}^{j+1}$ represent two adjacent imaginary-time slices, while the subscripts $l$ and $m$ denote the beads in the two slices. If we use $w_{\text{max}}$ to denote the cutoff in the summation over the integer vector $\mathbf{W}$, a direct summation of the above formula takes $O((w_{\text{max}})^d)$ time where $d$ is the dimension, since there are $O((w_{\text{max}})^d)$ number of terms inside the summation, which is inefficient. 
We rewrite Eq. (\ref{Wsum}) as
\begin{equation}
\begin{split}
A(\textbf {R}^j,\textbf {R}^{j+1})_{l,m}&=\sum_{\mathbf{W}}\Pi_{i=1}^d\exp\left(-\frac{1}{2\Delta\tau}\left(({R}_{l}^j)_i-( {R}^{j+1}_m)_i+(\mathbf{W}L)_i\right)^2\right)\\
&=\Pi_{i=1}^d\sum_{{W}_i}\exp\left(-\frac{1}{2\Delta\tau}\left(({R}^j_{l})_i-( {R}^{j+1}_m)_i+W_iL\right)^2\right).
\end{split}
\label{nsum}
\end{equation}
Here, $(R_l^j)_i$ denotes the spatial component of the $l$-th bead (in the $j$-th imaginary-time slice) along the $i$-th direction. So we evaluate the summation over each dimension and multiply the results together to obtain the final result. Now we are doing summation over each spatial dimension separately, with one summation taking $O(w_{\text{max}})$ time. There are $d$ summations to perform, so the algorithm scales as $O(d\times w_{\text{max}})$ in this way.

To calculate average energy we use thermodynamic estimator, it can be shown that the explicit expression for the estimator is
\begin{equation}
E=\frac{dN}{2\Delta\tau}+\sum_{j=1}^M\left[\langle -\mathrm{Tr} (A(\textbf{R}^j,\textbf{R}^{j+1})^{-1}\frac{\partial}{\partial\beta}A(\textbf{R}^j,\textbf{R}^{j+1}))\rangle+\langle \frac{V_{\text{int}}(\mathbf{R}^j)}{M}\rangle\right].
\end{equation}
Here, the operator $\mathrm{Tr}$ denotes the matrix trace. $A(\textbf{R}^j,\textbf{R}^{j+1})$ is a $N\times N$ matrix, while $A(\textbf{R}^j,\textbf{R}^{j+1})^{-1}$ is its inverse matrix. $\frac{\partial}{\partial \beta}$ acts as a partial derivative on each element of the matrix $A(\mathbf{R}^j, \mathbf{R}^{j+1})$. When we view $A(\mathbf{R}^j, \mathbf{R}^{j+1})$ as a matrix, it is clear that the subscripts $(l, m)$ correspond to the elements of this matrix.

Moreover, to calculate the thermodynamic energy estimator we also need the derivative of $A(\textbf {R}^j,\textbf {R}^{j+1})_{l,m}$ with respect to $\beta$:
\begin{equation}
\frac{\partial A(\textbf {R}^j,\textbf {R}^{j+1})_{l,m}}{\partial\beta}=\sum_{\mathbf{W}}\frac{1}{2\Delta\tau\beta}\left(\textbf {R}^j_{l}-\textbf {R}^{j+1}_m+\textbf{W}L\right)^2\exp\left(-\frac{1}{2\Delta\tau}\left(\textbf {R}^j_{l}-\textbf {R}^{j+1}_m+\textbf{W}L\right)^2\right).
\end{equation}
Using Eq. (\ref{nsum}) and 
\begin{equation}
\frac{\partial A(\textbf {R}^j,\textbf {R}^{j+1})_{l,m}}{\partial\beta}=A(\textbf {R}^j,\textbf {R}^{j+1})_{l,m}\frac{\partial\ln A(\textbf {R}^j,\textbf {R}^{j+1})_{l,m} }{\partial\beta},
\end{equation}
it is straightforward to have
\begin{equation}
\frac{\partial A(\textbf {R}^j,\textbf {R}^{j+1})_{l,m}}{\partial\beta}=A(\textbf {R}^j,\textbf {R}^{j+1})_{l,m}\sum_{i=1}^d\epsilon_i,
\label{Wsum2}
\end{equation}
where $\epsilon_i$ is defined as
\begin{equation}
\epsilon_i=\frac{\sum_{W_i}\frac{1}{2\Delta\tau\beta}\left(({R}_{l}^j)_i-( {R}^{j+1}_m)_i+W_iL\right)^2\exp\left(-\frac{1}{2\Delta\tau}\left(({R}_{l}^j)_i-( {R}^{j+1}_m)_i+W_iL\right)^2\right)}{\sum_{W_i}\exp\left(-\frac{1}{2\Delta\tau}\left(({R}_{l}^j)_i-( {R}^{j+1}_m)_i+W_iL\right)^2\right)}.
\end{equation}
As in Eq. (\ref{nsum}), evaluating Eq. (\ref{Wsum2}) only takes $O(d\times w_{\text{max}})$ time.

\section{Results}
\label{results}

\subsection{Spin-polarized uniform electron gas}

As a typical example of strongly quantum-degenerate dense uniform electron gas, we first consider the spin-polarized uniform electron gas at $r_s=0.5,\theta=0.0625$. For $N=4$, CPIMC yields the exact energy \cite{CPIMC-4}, which allows us to benchmark the pseudo-fermion method in small systems. From the definition of 
\begin{equation}
X(\beta,\lambda,M)=\frac{Z_F(\beta,\lambda,M)}{Z_{pf}(\beta,\lambda,M)},
\end{equation}
we may simulate $X(\beta,\lambda,M)$ by direct PIMC for small Fermi system, similar to the simulation of the sign factor $S(\beta,\lambda,M)$ (See Eq. (\ref{Sfactor})) given by the ratio between the usual expression of the bosonic partition function and the fermionic partition function \cite{Dornheim}. 

\begin{figure}[htbp]
\begin{center}
\includegraphics[width=0.7\textwidth]{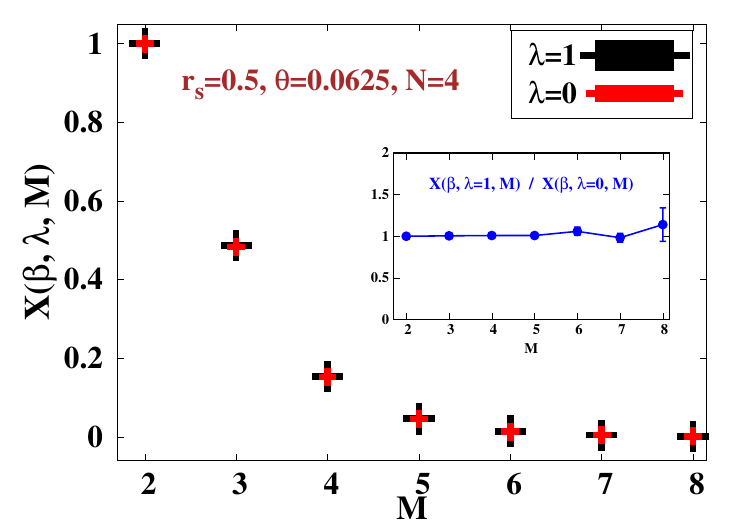}
\caption{\label{Xfactor} $X(\beta,\lambda,M)$ for $N=4$, $r_s=0.5$ and $\theta=0.0625$. Black crosses with error bars correspond to $\lambda=1$, while the red crosses correspond to $\lambda=0$. The inset displays the ratio $X(\beta,\lambda=1,M)/X(\beta,\lambda=0,M)$ for $M$ up to 8.}
\end{center}
\end{figure}

In the previous section (Sec. \ref{Mcana}), we pointed out that a key requirement for the validity of the pseudo-fermion method is the weak dependence of $X(\beta, \lambda, M)$ (or $\mathcal{O}(\beta,\lambda,M)$) on $\lambda$. In Fig.~\ref{Xfactor}, we used direct PIMC to simulate $X(\beta,\lambda,M)$ for $N=4, r_s=0.5$ at $\theta=0.0625$.
In Fig. \ref{Xfactor}, the black crosses with error bars represent the case of $\lambda=1$, while the red crosses represent $\lambda=0$. We observe that within the range of fluctuations, $X(\beta,\lambda,M)$ for $\lambda=0$ and $\lambda=1$ agree well, indicating that in this case $X(\beta,\lambda,M)$ indeed depends only very weakly on $\lambda$, thereby supporting the reliability of the pseudo-fermion method. In the inset, we present $\mathcal{O}(\beta,\lambda,M)=X(\beta,\lambda=1,M)/X(\beta,\lambda=0,M)$ for values up to $M=8$, demonstrating that $\mathcal{O}(\beta,\lambda,M)$ exhibits a very weak dependence on $\lambda$. For $M>8$, since both $X(\beta,\lambda=1,M)$ and $X(\beta,\lambda=0,M)$ drop below $10^{-3}$, numerical limitations preclude a reliable presentation of the data.
Building on the weak $\lambda$-dependence of $X(\beta, \lambda, M)$, we can further optimize $M$ to accurately determine the fermionic energy via the pseudo-fermion method free from sign-problem-induced instabilities.

\begin{figure}[htbp]
\begin{center}
\includegraphics[width=0.8\textwidth]{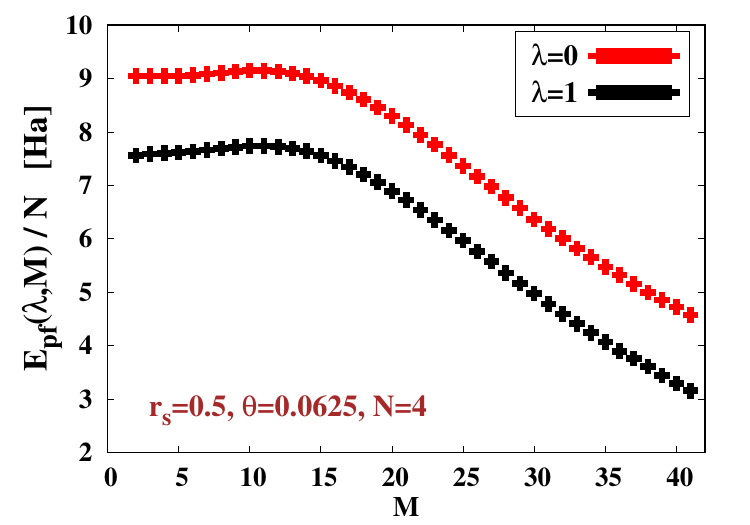}
\caption{\label{N4polarize1} For the case of $N=4$, $r_s=0.5$, and $\theta=0.0625$,
the red and black crosses with error bars represent the simulation results of the average energy per pseudo-fermion at different $M$ for $\lambda=0$ and $\lambda=1$, respectively.
}
\end{center}
\end{figure}

\begin{figure}[htbp]
\begin{center}
\includegraphics[width=0.8\textwidth]{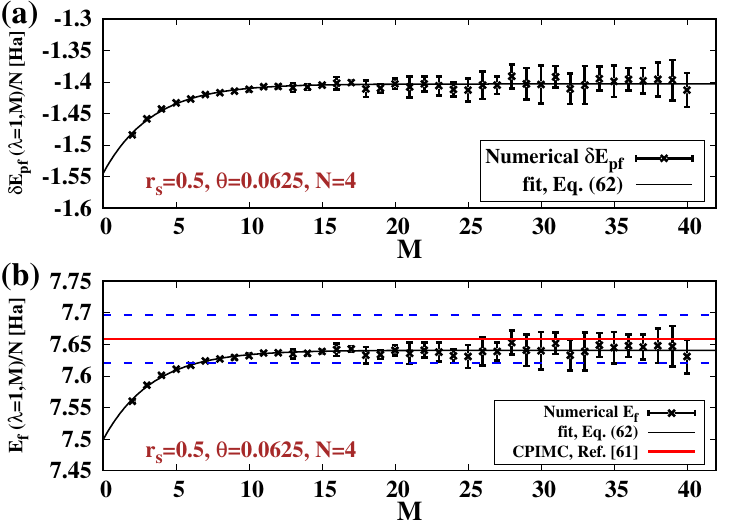}
\caption{\label{N4polarize2} For the case of $N=4$, $r_s=0.5$, and $\theta=0.0625$:
(a) The black crosses with error bars represent $\delta E_{pf}(\lambda=1,M)/N$ at different $M$ values. The black curve shows the fit using the function $f(M)$.
(b) The black crosses with error bars represent the inferred fermion energy at different $M$ values, and the black curve is the corresponding fit using $f(M)$. The red line indicates the exact result from CPIMC \cite{CPIMC-4}, and the two blue dashed lines correspond to a $0.5\%$ upward and downward shift of the red line.}
\end{center}
\end{figure}

For $N=4, r_s=0.5, \theta=0.0625$, we now begin to use the pseudo-fermion method shown in Sec. \ref{Mcana} to infer the fermionic energy. In Fig.~\ref{N4polarize1}, the red and black crosses with error bars represent the pseudo-fermion simulation results for the average energy per particle at $\lambda=0$ and $\lambda=1$, respectively, as a function of $M$. From the results in Fig.~\ref{N4polarize1}, in Fig.~\ref{N4polarize2}(a), the black crosses with error bars represent $\delta E_{pf}(\beta,\lambda,M)$. At $M=2$, one does not expect an accurate inference of the fermionic energy. From Fig.~\ref{N4polarize2}(a), we see that $\delta E_{pf}(\beta,\lambda,M)$ rises rapidly from $M=2$ and then enters a plateau region with fluctuations. The fluctuations of $\delta E_{pf}(\beta,\lambda,M)$ increase with $M$, possibly because we did not use the virial estimator \cite{virial} in this work. For fictitious identical particles and boltzmannons, the virial estimator \cite{Xiong-qua} can be incorporated in the usual way. For pseudo-fermions and fermions in PIMC simulations based on the fermionic propagator, how to incorporate a virial estimator remains an open question for future research.

To account for the plateau behavior and fluctuations of $\delta E_{pf}(\beta,\lambda,M)$, we fit it with the function
\begin{equation}
f(M)=a+be^{-c M}.
\label{aexp}
\end{equation}
In Fig.~\ref{N4polarize2}(a), the black curve shows the fitting result. Since $f(M)$ becomes nearly flat for large $M$, in this work we take $a$ as the extrapolated value of $\delta E_{pf}$ to infer the fermionic energy.

$\delta E_{pf}$ itself is not the fermionic energy; one must add the ideal Fermi gas energy to infer the fermionic energy. In Fig.~\ref{N4polarize2}(b), we show the inferred fermionic energy for different $M$. The black curve shows the fit using $f(M)$. The red line represents the exact CPIMC result \cite{CPIMC-4}, and the two blue dashed lines represent $\pm 0.5\%$ relative deviations from the red line. The pseudo-fermion estimate differs from the exact fermionic energy by only $0.24\%$. At $M=2$, where no sign problem occurs, the relative deviation is $1.3\%$. This comparison indicates that the pseudo-fermion method, while avoiding the sign problem in simulations, can significantly improve the accuracy of fermionic energy estimates.

In Fig. \ref{N4polarize2}, we observe that the plateau region is remarkably broad, remaining essentially constant within the range of fluctuations from $M=15$ to $M=40$. This implies that the bias introduced by the Suzuki-Trotter decomposition is negligible, and the total error is dominated by the omission of the dependence of $\delta E_{\mathcal{O}}$ on $\lambda$. Of course, if the plateau were much narrower, the systematic bias resulting from the finite-$M$ Suzuki-Trotter expansion might play a more significant role.

\begin{figure}[htbp]
\begin{center}
\includegraphics[width=0.48\textwidth]{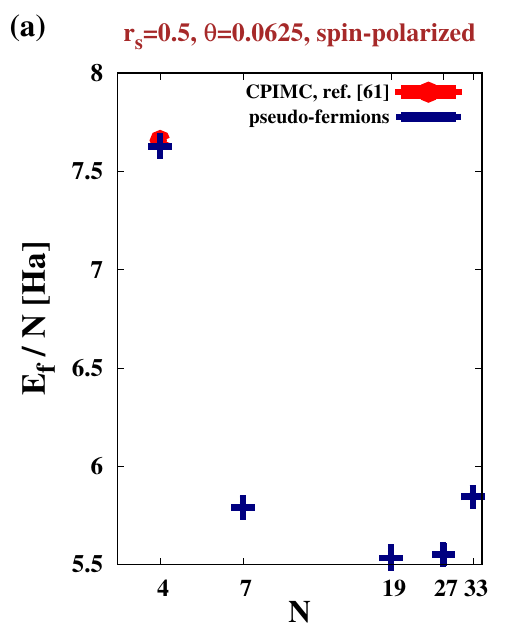}
\includegraphics[width=0.48\textwidth]{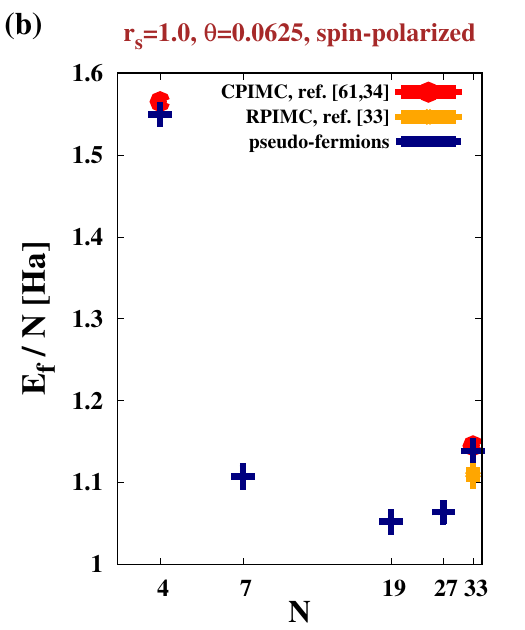}
\caption{\label{PolarizedN} For the case of $r_s=0.5$ and $\theta=0.0625$, (a) shows the energy per fermion for different particle numbers as obtained by the pseudo-fermion method (blue crosses with error bars) and CPIMC \cite{CPIMC-4} (red dot).
For the case of $r_s=1.0$ and $\theta=0.0625$, (b) shows the energy per fermion for different particle numbers as obtained by the pseudo-fermion method (blue crosses with error bars), CPIMC \cite{CPIMC,CPIMC-4} (red dots), and RPIMC \cite{Brown} (yellow dot with error bar).}
\end{center}
\end{figure}

Once the pseudo-fermion results agree closely with CPIMC for $N=4$, we can extend the method to larger $N$. In Fig.~\ref{PolarizedN}(a), for $r_s=0.5,\theta=0.0625$, the blue crosses with error bars represent the pseudo-fermion results for $N=4,7,19,27,33$, while the red dot shows CPIMC result \cite{CPIMC-4} for $N=4$. It is worth emphasizing that RPIMC simulation \cite{Brown} fails to provide reliable results at $r_s=0.5$, as shown in Ref. \cite{CPIMC}. In Ref. \cite{CPIMC}, no CPIMC simulation result is provided for $r_s = 0.5$, which offers an opportunity for an independent verification of the pseudo-fermion method using CPIMC in the future. 

In Fig.~\ref{PolarizedN}(b), we show results for $r_s=1,\theta=0.0625$. The blue crosses with error bars are the results based on pseudo-fermion method for $N=4,7,19,27,33$, while the red dots and yellow dot with error bar represent CPIMC \cite{CPIMC,CPIMC-4} and RPIMC \cite{Brown} results, respectively. For $N=33$, the result based on pseudo-fermion method agrees closely with CPIMC, while RPIMC deviates significantly. For $N=33$, the pseudo-fermion and CPIMC results \cite{CPIMC} agree within $0.6\%$, well within the fluctuations of the result of pseudo-fermion method.  
In both the $r_s=0.5$ and $r_s=1$ cases, the results based on pseudo-fermion method are slightly lower than the exact CPIMC results, which is due to our neglect of $\delta E_{\mathcal{O}}(\beta,\lambda,M)$.

In Fig.~\ref{PolarizedN}, we also observe a marked decrease in $E_f/N$ when going from $N=4$ to $N=7$. This is because for periodic boundary conditions, $N=7,19,27,33$ correspond to closed-shell fillings, whereas $N=4$ does not. Even for closed shells, $E_f/N$ does not behave monotonically with $N$. Indeed, the same non-monotonic behavior is found in the noninteracting case.

\begin{figure}[htbp]
\begin{center}
\includegraphics[width=0.48\textwidth]{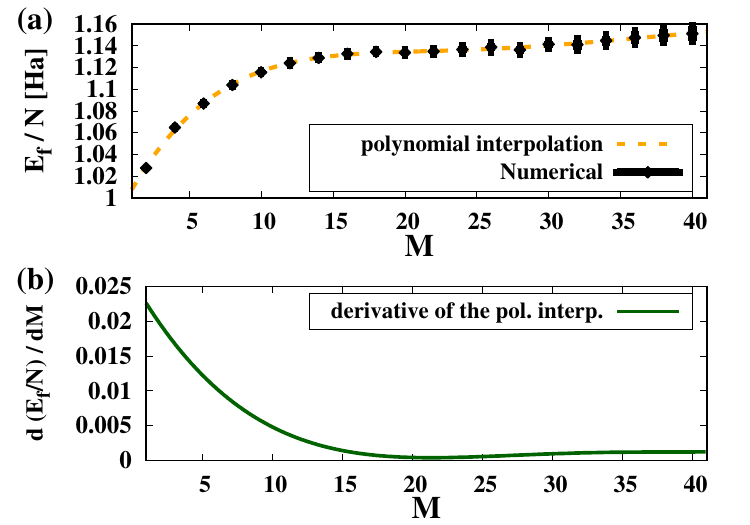}
\includegraphics[width=0.48\textwidth]{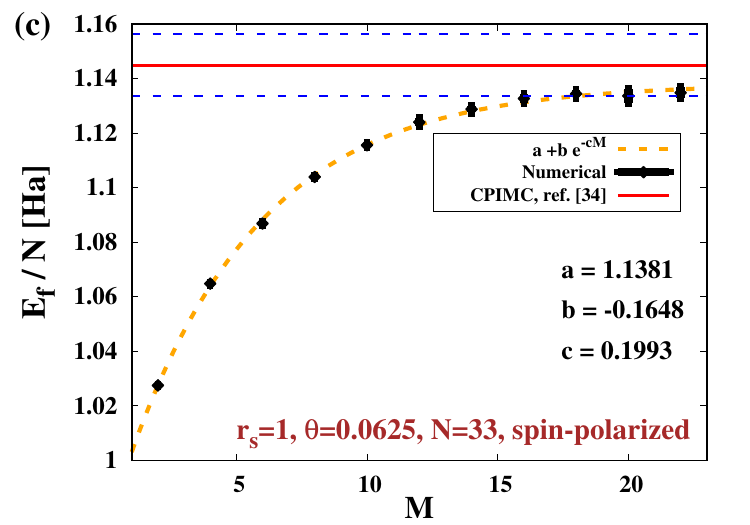}
\caption{\label{rs1_high_acc} Estimated fermion energy as a function of $M$ for $r_s=1$ and $\theta=0.0625$. Panel (a): fermionic energy as a function of $M$, interpolated using a 5th order polynomial. Panel (b): derivative of the polynomial function in (a). Panel (c): extrapolated fermionic energy (dashed orange line). The red line indicates the exact result from CPIMC \cite{CPIMC}, and the two blue dashed lines correspond to a $1\%$ upward and downward shift of the red line.}
\end{center}
\end{figure}

In Fig. \ref{rs1_high_acc}, we provide the details for the extrapolation of the fermionic energy for the case where $N=33$, $r_s=1$, and $\theta=0.0625$.
In Fig. \ref{rs1_high_acc}(a), we show the relationship between $E_f/N = \left(E_f(\lambda=0) + \delta E_f(\lambda=1, M)\right)/N$ and $M$. Data have been obtained using $10$ independent simulations each of which collected $4 \times 10^4$ highly de-correlated configurations. 
The yellow dashed line in Fig. \ref{rs1_high_acc}(a) represents the result of a polynomial fit. Fig. \ref{rs1_high_acc}(b) displays the derivative of the polynomial fitting function with respect to $M$, clearly indicating the presence of a minimum at $M=22$. 

In Fig. \ref{rs1_high_acc}(c), we show the extrapolated fermionic energy obtained using an exponential fit. We excluded data points where $M>22$ so that the exponential fit could better preserve the morphology of the flat region. We also note that the result of the exponential fit is essentially consistent with the extrapolation based directly on the $M=22$ data point. The reason for employing the exponential fit is to better suppress fluctuations arising from the numerical simulation. Fig. \ref{rs1_high_acc} indicates that for the case of $r_s=1$, the pseudo-fermion method can self-consistently extrapolate the fermionic energy based on the flat region shown in the simulation. In the systematic study that follows, we applied the same analysis for every $r_s$ value to ensure the flat region was appropriately considered.

We stress that for a finite $M$, the Suzuki-Trotter decomposition introduces systematic errors. However, when a larger $M$ is employed, it may exceed the flat region, leaving us without a clear rule for determining the optimal $M$ required to infer the fermion energy. To ensure the reliability of the specific operational procedures of the pseudo-fermion method presented here, we suggest that when applying it to a specific system, one should first perform simulations on small-scale systems with rigorous benchmarks to assess the validity before extending it to larger-scale fermionic systems.

\begin{figure}[htbp]
\begin{center}
\includegraphics[width=0.8\textwidth]{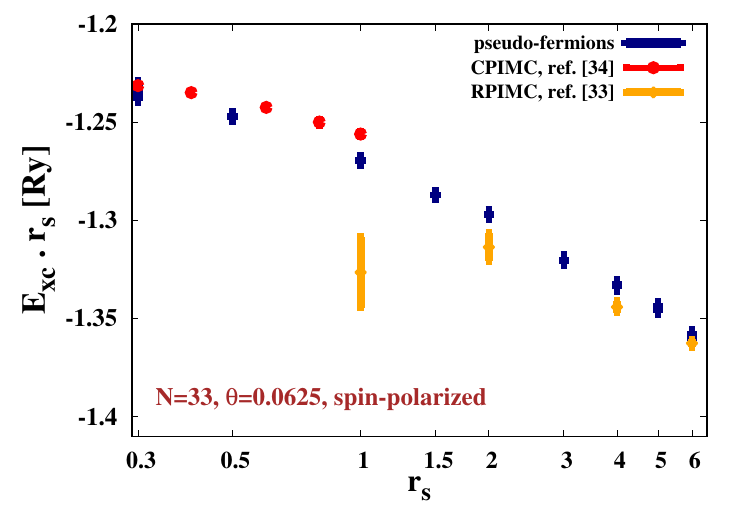}
\caption{\label{Exc_tomm} Exchange-correlation energy for $(N=33,\theta=0.0625)$ as a function of $r_s$ in the fully polarized case. Blue points correspond to the estimated value using pseudo-fermions, red points are CPIMC data taken from Ref. \onlinecite{CPIMC}, while orange points are the RPIMC data from Ref. \onlinecite{Brown}.
}
\end{center}
\end{figure}

In Fig.~\ref{Exc_tomm}, we show the exchange-correlation energy
\begin{equation}
E_{XC}=\frac{E_f(\lambda=1)-E_f(\lambda=0)}{N}
\end{equation}
for different $r_s$. For ease of comparison, the definition of the exchange–correlation energy used here is the same as that in Ref. \cite{CPIMC}. The blue crosses with error bars are the results based on pseudo-fermion method, the red dots are the results of CPIMC \cite{CPIMC} while the yellow dots with errors correspond to the results of RPIMC \cite{Brown}. 
We found that the results of pseudo-fermion method are clearly closer to CPIMC than RPIMC for $r_s=1$. Since both pseudo-fermion method and RPIMC avoid the sign problem during simulations, this comparison highlights the value of the pseudo-fermion method. It is worth pointing out that PB-PIMC \cite{Groth}, based on the fermionic propagator and high-order Suzuki-Trotter decompositions \cite{PB}, cannot simulate as many as $N=33$ polarized electrons in the strongly degenerate dense UEG considered here due to the severe fermion sign problem.

It is significant that our approach provides reliable results, particularly in the challenging intermediate-density region ($r_s \approx 1$), while retaining satisfactory accuracy at both high and low densities. Currently, no other single method has demonstrated comparable performance across this entire density range.

\subsection{Further discussions}

It must be emphasized that the pseudo-fermion method does not represent an ultimate solution to the fermion sign problem. Fortunately, for systems such as quantum dots \cite{pseudo} and the uniform electron gas discussed here, we find that the pseudo-fermion method offers an effective alternative for simulating fermionic systems. To ensure reliability in uniform systems, we believe that a robust strategy involves first comparing results with rigorous benchmarks in small-scale systems—where the efficacy of the pseudo-fermion method can be clearly observed—before extending the application to larger systems. Within this framework, the effective implementation of the pseudo-fermion method relies on empirical insights derived from phenomena observed during the simulation process. In the following, we provide several heuristic discussions regarding the application of the pseudo-fermion method to the uniform electron gas.

In the limit $r_s \rightarrow 0$, as the kinetic energy contribution far outweighs that of the Coulomb interaction, we have $\frac{\delta E_{pf}(\beta, \lambda; \text{flat})}{E_f(\beta, \lambda)} \rightarrow 0$. Within the plateau region, our previous analysis suggests the conjecture that $\delta E_\mathcal{O}(\beta, \lambda; \text{flat}) \ll \delta E_{pf}(\beta, \lambda; \text{flat})$. According to Eq. (\ref{deltaf}), even under the more conservative assumption that $\delta E_\mathcal{O} \sim \delta E_{pf}$, it follows that $\Delta_f(\beta, \lambda; \text{flat}) \rightarrow 0$ as $r_s \rightarrow 0$. This  behavior provides a physical justification for the validity of the pseudo-fermion method at small $r_s$. While a rigorous mathematical proof for the relationship between $\delta E_\mathcal{O}$ and $\delta E_{pf}$ is not provided here, for $r_s\rightarrow 0$, we believe this constitutes a promising direction for future systematic investigation.

The selection of Eq. (\ref{aexp}) for our fitting procedure is empirically motivated by numerical observations and cross-validations with exact benchmarks in small systems. Across a wide range of parameters for both small and large-scale systems (\textit{e.g.}, $N=33$ polarized electrons), we consistently observe that $\delta E_{pf}$ exhibits a rapid initial growth with $M$ followed by an asymptotic approach to a well-defined plateau. Eq. (\ref{aexp}) offers a simple yet physically grounded model that accurately captures this characteristic convergence behavior. Notably, this choice is not unique; as demonstrated in Fig. \ref{rs1_high_acc}, a polynomial fit also yields consistent results. For all data presented in Fig. \ref{Exc_tomm}, we performed comparative analyses using both Eq. (\ref{aexp}) and polynomial functions, finding no statistically significant discrepancies in the final outcomes. For future applications of the pseudo-fermion method to other fermionic systems, we recommend selecting a fitting functional form that best reflects the specific behavior of $\delta E_{pf}$ observed in those systems.

The plateau correspondence assumption is physically grounded in the coupled nature of path discretization and interaction representation within the path-integral framework. Specifically, the number of imaginary-time slices $M$ defines the resolution of the quantum paths; when the simulation enters the flat region with respect to $M$, it signifies that the discretization $\Delta\tau = \beta/M$ is sufficiently small to yield a converged representation of the kinetic density matrix. Since the interaction $\lambda$ is effectively smeared over these resolved paths, the stability of the underlying path structure (the $M$-plateau) naturally implies a corresponding stabilization of the interaction-induced correction $\delta E_{\mathcal{O}}$. In this convergent regime, the path configurations are dominated by the free-fermion propagator's exchange anti-symmetry, and the interaction $\lambda$ merely modulates the weights without altering the fundamental sign structure. Consequently, the sensitivity of $\delta E_{\mathcal{O}}$ to $\lambda$ (\textit{i.e.}, $\partial \delta E_{\mathcal{O}} / \partial \lambda$) is expected to be minimal once the $M$-dependence has plateaued. This coupled convergence is empirically validated by our cross-checks with exact CPIMC benchmarks, where the results extracted from the $M$-plateau consistently yield high-precision agreement across a wide range of $r_s$ values, confirming that the $M$-direction stability is a reliable proxy for the overall accuracy of the pseudo-fermion correction.

While a rigorous mathematical proof remains elusive, our \textit{ansatz} is robustly supported by the quantitative agreement with the established benchmarks. As shown in Fig. \ref{Exc_tomm}, the results obtained by taking the data from the $M$-plateau consistently match the reliable benchmarks across different $r_s$ values. If the dependence of $ \delta E_{\mathcal{O}}$ on $\lambda$ were strong or erratic in this region, we would expect to see a failure of the method as $r_s$ (and thus the relative strength of $\lambda$) increases. The fact that the error remains small and systematic across the entire range validates that the $M$-direction plateau is a reliable indicator of the overall convergence of the pseudo-fermion method.

\section{Summary and Discussion}

\label{summary}

In summary, the fermion sign problem arises from the antisymmetric nature of the fermionic propagator. In the pseudo-fermion approach\cite{pseudo}, a different propagator is used in path integral formalism which does not lead to sign problem. Even if the new pseudo-fermion partition function is not equivalent to that of the original fermionic system, we have chance to recover properties of the true fermionic system through correction relative to the noninteracting system, as supported by general mathematical analysis. 

In this work, we have applied the pseudo-fermion method to the strongly degenerate uniform electron gas and found excellent agreement with CPIMC \cite{CPIMC,CPIMC-4} for the density parameter it can be applied. Since CPIMC still suffers from the sign problem, the pseudo-fermion method offers the potential for simulating much larger fermionic systems with higher efficiency. 
For dense uniform electron gas with $r_s<2$, RPIMC becomes unreliable, whereas the pseudo-fermion method remains applicable.
Another advantage of the pseudo-fermion method over RPIMC is that it does not require the fixed-node approximation based on the finite-temperature density matrix of free fermions or other sophisticated schemes. This makes the method simpler and potentially more broadly applicable for first-principles simulations. For instance, in dense hydrogen, the pseudo-fermion method—free of fixed-node constraints—offers the possibility of treating both protons and electrons on equal footing within the PIMC framework. 
For dense hydrogen and beryllium \cite{DornheimA1,DornheimA2,DornheimA3}, the fictitious identical particle method with isothermal $\xi$-extrapolation has already been shown to work in weakly and moderately degenerate regimes when nuclei and electrons are treated together in PIMC. However, in the case of strong quantum degeneracy, the isothermal $\xi$-extrapolation no longer holds. One reason for this is that, at zero temperature, the fictitious identical-particle partition function exhibits elegant and universal Lee–Yang zeros \cite{Li,Fan}. In contrast, for the pseudo-fermion partition function, such Lee–Yang zeros associated with fictitious identical particles do not exist at zero temperature. Therefore, for the pseudo-fermion partition function, the fermionic energy at a given temperature can be inferred by tuning the number of imaginary-time slices $M$ (isothermal $M$-extrapolation) in the strongly degenerate uniform electron gas, making the approach promising for applications to dense hydrogen and beryllium.

The purpose of this work has been to demonstrate the application of the pseudo-fermion method to the uniform electron gas, rather than to optimize precision and efficiency. We believe that future developments—such as incorporating worm algorithms \cite{Burov1,Burov2,PB}, high-order Suzuki-Trotter decompositions \cite{ChinF,Brualla,PB}, and pair approximations \cite{CeperleyRMP,Pollock,Militzer,Bohme}—will further enhance both the efficiency and accuracy of the pseudo-fermion method when applied to the uniform electron gas.
As shown in this work and previous work on quantum dots\cite{pseudo}, the pseudo-fermion method works exceptionally well at the low temperature (strongly degenerate) regime. When pseudo-fermion method is combined with traditional diffusion quantum Monte Carlo \cite{CeperleyD,Foulkes} and path-integral diffusion Monte Carlo \cite{Hetenyi,Sarsa,Cuervo,Rota,Yan}, in the future there is chance to extend the applications to molecular systems relevant to quantum chemistry, high-pressure hydrogen \cite{Monacelli} at low temperatures and cold atoms \cite{Lode}. 

\section*{Acknowledgments}
Y. Xiong gratefully acknowledges the support of the Hubei Provincial Young Top-Talent Program for this work.

\appendix
\renewcommand{\thefigure}{A\arabic{figure}}
\setcounter{figure}{0}

\section{A derivation of the expression for the partition function of fermions}
\label{appendix}

The derivation of how to express the fermionic partition function in terms of the fermion propagator has been scarcely presented in past literature; therefore, we provide some of the details here. For the sake of convenience and simplicity, we do not consider the additional complications introduced by the periodic box here. The extension to the case of a periodic box is straightforward.

For $N$ distinguishable particles, we can define the following coordinate basis
\begin{equation}
\left|\textbf R\right>_D=\left|\textbf r_1,\textbf r_2,\cdots,\textbf r_N\right>_D.
\end{equation}
We emphasize here that in the state $\left|\mathbf{r}_1, \mathbf{r}_2, \cdots, \mathbf{r}_N\right>_D$, there exists an ordering from the first particle to the $N$-th particle. For another coordinate basis
\begin{equation}
\left|\textbf R'\right>_D=\left|\textbf r'_1,\textbf r'_2,\cdots,\textbf r'_N\right>_D,
\end{equation}
we have
\begin{equation}
_D\left<\textbf R\right|\left.\textbf R'\right>_D=\delta(\textbf r_1-\textbf r'_1)\delta(\textbf r_2-\textbf r'_2)\cdots \delta(\textbf r_N-\textbf r'_N).
\end{equation}
In obtaining the inner product of the above expressions, we of course need to proceed sequentially according to the order of the particles.

Starting from $\left|\textbf R\right>_D$, we can construct the following fermionic coordinate basis $\left|\textbf R\right>_F$:
\begin{equation}
\left|\textbf R\right>_F=\frac{1}{\sqrt{N!}}\sum_P(-1)^{N_P} \left|P\{\textbf r_1,\textbf r_2,\cdots,\textbf r_N\}\right>_D.
\end{equation}
Here, \( P \) represents the permutation operator acting on the coordinates, and \( N_P \) denotes the minimal number of pairwise exchanges required to restore the original order of the coordinates under permutation \( P \). It should be noted that in this convention, the permutation operator $P$ above permutes the coordinates themselves without changing the ordering from the first particle to the $N$-th particle. For example, for the permuted state $\left|\textbf r_2, \textbf r_1, \cdots, \textbf r_N\right>_D$, it represents that the first particle has coordinate $\textbf r_2$ and the second particle has coordinate $\textbf r_1$. Obviously, for the fermionic state $\left|\textbf R\right>_F$ defined above, we have $\left|\textbf R\right>_F \neq 0$ only if all $\textbf r_j$ are distinct. 

Based on the above definition of the fermionic coordinate basis, it is straightforward to show that this basis satisfies the following antisymmetry property:
\begin{equation}
\left|\textbf R\right>_F=(-1)^{N_P}\left|P\textbf R\right>_F.
\label{antiP}
\end{equation}

Let us consider the following inner product:
\begin{equation}
_F\left<\textbf R\right|\left.\textbf R'\right>_F=\frac{1}{N!}\sum_P\sum_{P'}(-1)^{N_P}(-1)^{N_{P'}}
~_D\left<P\{\textbf r_1,\textbf r_2,\cdots,\textbf r_N\}\right |\left. P'\{\textbf r'_1,\textbf r'_2,\cdots,\textbf r'_N\}\right>_D.
\end{equation}
For the above inner product, we consider the case where $\textbf r_1 \to \textbf r'_1, \textbf r_2 \to \textbf r'_2, \ldots, \textbf r_N \to \textbf r'_N$. In the above expression, we immediately notice that for a given $P$, $P'$ must be equal to $P$ for the term to be nonzero. Thus, we obtain
\begin{equation}
_F\left<\textbf R\right|\left.\textbf R'\right>_F=\delta(\textbf r_1-\textbf r'_1)\delta(\textbf r_2-\textbf r'_2)\cdots \delta(\textbf r_N-\textbf r'_N).
\end{equation}
For the above relation, of course, we also require that all $\textbf r_j$ are distinct.

For the fermionic partition function, we have
\begin{equation}
Z_F(\beta)=Tr (e^{-\beta\hat H})=\frac{1}{N!}\int d\textbf R\sum_P\sum_{P'}(-1)^{N_P}(-1)^{N_{P'}}~_D\left<P\textbf R\right|e^{-\beta\hat H}\left |P'\textbf R\right>_D.
\end{equation}
It is worth noting that the Hamiltonian operator $\hat H$ is also invariant under the exchange of any two particles for uniform electron gas.

For the double sum over $P$, $P'$ in the above partition function, let us consider the case with $P'$ fixed:
\begin{equation}
Z_F(P',\beta)=\frac{1}{N!}\int d\textbf R\sum_P(-1)^{N_P}(-1)^{N_{P'}}~_D\left<P\textbf R\right|e^{-\beta\hat H}\left |P'\textbf R\right>_D.
\label{ZP'}
\end{equation}
Based on Eq. (\ref{antiP}), we have
\begin{equation}
\sum_P(-1)^{N_P}~_D\left<P\textbf R\right|=(-1)^{N_{P'}}\sum_P(-1)^{N_P}~_D\left<P (P'\textbf R)\right|.
\end{equation}
Substituting this relation into Eq. (\ref{ZP'}), we have
\begin{equation}
\begin{split}
Z_F(P',\beta)=&
\frac{1}{N!}\int d\textbf R\sum_P(-1)^{N_P}~_D\left<P(P'\textbf R)\right|e^{-\beta P'\hat H}\left |P'\textbf R\right>_D.
\end{split}
\end{equation}
In the above derivation, we also used $P'\hat H = \hat H$. In the present expression, when $P'$ acts on $\textbf R$ to perform a permutation, it does not introduce an additional sign factor. 
Since $\hat H$ is invariant under any permutation $P'$, all the $P'$ in the above expression can be safely omitted, yielding
\begin{equation}
\begin{split}
Z_F(P',\beta)=&
\frac{1}{N!}\int d\textbf R\sum_P(-1)^{N_P}~_D\left<P\textbf R\right|e^{-\beta \hat H}\left |\textbf R\right>_D.
\end{split}
\end{equation}
The above derivation shows that $Z_F(P',\beta)$ is independent of $P'$. From this, we then have
\begin{equation}
Z_F(\beta)=\int d\textbf R\sum_P(-1)^{N_P}~_D\left<P\textbf R\right|e^{-\beta\hat H}\left |\textbf R\right>_D.
\end{equation}
In the conventional definition of the fermion partition function, there is an additional factor of $1/N!$, but the above derivation shows that this extra factor is not necessary.

We decompose $e^{-\beta \hat H}$ into $e^{-\Delta \tau \hat H} \cdots e^{-\Delta \tau \hat H}$. From a mathematical perspective, we can insert the following unit operator for distinguishable particles to obtain the currently popular expression for the fermionic partition function:
\begin{equation}
\hat I_D=\int d\textbf R~_D\left|\textbf R\right>\left<\textbf R\right|_D.
\end{equation}
This procedure is justified because $\hat I_D$ is complete in the Hilbert space, and it can be easily shown that $\hat I_D^2 = \hat I_D$.

In this work, we adopt a different form of the fermionic partition function. To fully derive the fermionic partition function constructed from the fermionic propagator, we consider the following imaginary-time propagator:
\begin{equation}
\rho_F(\textbf R,\textbf R';\Delta\tau)=~_F\left<\textbf R\right|e^{-\Delta\tau\hat H}\left|\textbf R'\right>_F=\frac{1}{N!}\sum_P\sum_{P'}(-1)^{N_P}(-1)^{N_{P'}}~_D\left<P\textbf R\right|e^{-\Delta\tau\hat H}\left|P'\textbf R'\right>_D.
\end{equation}
Similar to the previous analysis, we consider a particular $P'$ in the above double summation.
\begin{equation}
\rho_F(\textbf R,\textbf R';P',\Delta\tau)=\frac{1}{N!}\sum_P(-1)^{N_P}(-1)^{N_{P'}}~_D\left<P\textbf R\right|e^{-\Delta\tau P'\hat H}\left|P'\textbf R'\right>_D.
\end{equation}
In the above expression, we have already replaced $\hat H$ with $P'\hat H$. Based on Eq. (\ref{antiP}), we have
\begin{equation}
\rho_F(\textbf R,\textbf R';P',\Delta\tau)=\frac{1}{N!}\sum_P(-1)^{N_P}~_D\left<P(P'\textbf R)\right|e^{-\Delta\tau P'\hat H}\left|P'\textbf R'\right>_D.
\end{equation}
Since $\hat H$ is invariant under the action of the permutation $P'$, $\rho_F(\mathbf R, \mathbf R';P',\Delta\tau)$ is independent of $P'$. Therefore, we have
\begin{equation}
\rho_F(\textbf R,\textbf R';\Delta\tau)=\sum_P(-1)^{N_P}~_D\left<P\textbf R\right|e^{-\Delta\tau\hat H}\left|\textbf R'\right>_D.
\end{equation}
This expression corresponds to the fermion propagator between adjacent imaginary-time slices used in the main text. The summation
$\sum_P (-1)^{N_P}$
allows $\rho_F(\mathbf R, \mathbf R';\Delta\tau)$ to be expressed in terms of a determinant.

\end{document}